\documentclass[english,prl, superscriptaddress, longbibliography, onecolumn, notitlepage]{revtex4-1}
\usepackage[T1]{fontenc}
\usepackage[latin9]{inputenc}
\usepackage{float}
\usepackage{bm}
\usepackage{amsmath}
\usepackage{graphicx}

\makeatletter
\usepackage{amsmath}
 
\usepackage[normalem]{ulem}
\usepackage[colorlinks=true,linkcolor=MidnightBlue,urlcolor=MidnightBlue,citecolor=MidnightBlue,anchorcolor=MidnightBlue]{hyperref}\usepackage[dvipsnames]{xcolor}

\usepackage{chngcntr}
\usepackage{listings}
\usepackage{xcolor}
 
\definecolor{codegreen}{rgb}{0,0.6,0}
\definecolor{codegray}{rgb}{0.5,0.5,0.5}
\definecolor{codepurple}{rgb}{0.58,0,0.82}
\definecolor{backcolour}{rgb}{0.98,0.98,0.98}
 
\lstdefinestyle{mystyle}{
    backgroundcolor=\color{backcolour},   
    commentstyle=\color{codegreen},
    keywordstyle=\color{magenta},
    numberstyle=\tiny\color{codegray},
    stringstyle=\color{codepurple},
    basicstyle=\linespread{0.8}\footnotesize\ttfamily
    breakatwhitespace=false,         
    breaklines=true,                 
    captionpos=b,                    
    keepspaces=true,                 
    numbersep=5pt,                  
    showspaces=false,                
    showstringspaces=false,
    showtabs=false,                  
    tabsize=2
}
 
 \newcommand*{\bfrac}[2]{\genfrac{[}{]}{0pt}{}{#1}{#2}}

\makeatother

\usepackage{babel}
\usepackage{listings}

\begin{document}
\title{PyStokes: phoresis and Stokesian hydrodynamics in Python}
\begin{abstract}
We present a modular Python library for computing many-body hydrodynamic
and phoretic interactions between spherical active particles in suspension,
when these are given by solutions of the Stokes and Laplace equations.
Underpinning the library is a grid-free methodology that combines
dimensionality reduction, spectral expansion, and Ritz-Galerkin discretization,
thereby reducing the computation to the solution of a linear system.
The system can be solved analytically as a series expansion or numerically
at a cost quadratic in the number of particles. Suspension-scale quantities
like fluid flow, entropy production, and rheological response are
obtained at a small additional cost. The library is agnostic to boundary
conditions and includes, amongst others, confinement by plane walls
or liquid-liquid interfaces. The use of the library is demonstrated
with six fully coded examples simulating active phenomena of current
experimental interest. 
\end{abstract}
\author{Rajesh Singh}
\email{rs2004@cam.ac.uk}

\affiliation{DAMTP, Centre for Mathematical Sciences, University of Cambridge,
Wilberforce Road, Cambridge CB3 0WA, UK}
\author{R. Adhikari}
\email{ra413@cam.ac.uk, rjoy@imsc.res.in,}

\affiliation{DAMTP, Centre for Mathematical Sciences, University of Cambridge,
Wilberforce Road, Cambridge CB3 0WA, UK}
\affiliation{The Institute of Mathematical Sciences-HBNI, CIT Campus, Chennai 600113,
India}
\maketitle

\section{Introduction}

PyStokes is a Python library for studying phoretic and hydrodynamic
interactions between spherical particles when these interactions can
be described by the solutions of, respectively, the Laplace and Stokes
equations. The library has been specifically designed for studying
these interactions in suspensions of active particles, which are distinguished
by their ability to produce flow, and thus motion, in the absence
of external forces or torques. Such particles are endowed with a mechanism
to produce hydrodynamic flow in a thin interfacial layer, which may
be due to the motion of cilia, as in microorganisms \citep{brennen1977}
or osmotic flows of various kinds in response to spontaneously generated
gradients of phoretic fields \citep{ebbens2010pursuit}. The latter,
often called autophoresis, is a generalisation of well- known phoretic
phenomena including, inter alia, electrophoresis (electric field),
diffusiophoresis (chemical field) and thermophoresis (temperature
field) that occur in response to externally imposed gradients of phoretic
fields \citep{anderson1989colloid}.

Hydrodynamic and phoretic interactions between ``active particles''
in a viscous fluid are central to the understanding of their collective
dynamics \citep{ebbens2010pursuit,zhang2017active}. Under experimentally
relevant conditions, the motion of the fluid is governed by the Stokes
equation and that of the phoretic field, if one is present, by the
Laplace equation. The ``activity'' appears in these equations as
boundary conditions on the particle surfaces that prescribe the slip
velocity in the Stokes equation and flux of the phoretic field in
the Laplace equation. The slip velocity and the phoretic flux are
related by a linear constitutive law that can be derived from a detailed
analysis of the boundary layer physics \citep{anderson1989colloid}.
The Stokes and Laplace equations are coupled by this linear constitutive
law only at the particle boundaries. The linearity of the governing
equations and of the coupling boundary conditions allows for a formally
exact solution of the problem of determining the force per unit area
on the particle surfaces. This formally exact solution can be approximated
to any desired degree of accuracy by a truncated series expansion
in a complete basis of functions on the particle boundaries. This,
in turn, leads to an efficient and accurate numerical method for computing
hydrodynamic and phoretic interactions between active particles \citep{singh2015many,singh2018generalized,singh2019competing}. 

The principal features that set this method apart are (a) the restriction
of independent fluid and phoretic degrees of freedom to the particle
boundaries (b) the freedom from grids, both in the bulk of the fluid
and on the particle boundaries and (c) the ability to handle, within
the same numerical framework, a wide variety of geometries and boundary
conditions, including unbounded volumes, volumes bounded by plane
walls or interfaces, periodic volumes and, indeed, any geometry-boundary
condition combination for which the Green's functions of the governing
equations are simply evaluated.

The purpose of this article is to demonstrate the power of the numerical
method, as implemented in Python library, through six fully coded
examples that simulate experimental phenomena. Our software implementation
uses a polylgot programming approach that combines the readability
of Python with the speed of Cython and retains the advantages of a
high-level, dynamically typed, interpreted language without sacrificing
performance. 

Our presentation is in the style of literate programming and draws
inspiration from similar articles by Weideman and Reddy \citep{weideman2000matlab},
Higham \citep{higham2001algorithmic}, and Trefethen \citep{trefethen2000spectral}.
The article is best read alongside installing the library and executing
the example codes. The library freely is available on GitHub at \href{https://github.com/rajeshrinet/pystokes}{https://github.com/rajeshrinet/pystokes},
\ where detailed installation instructions can also be found. A subset
of the library features is available as a Binder file which requires
no installation. All software is released under the MIT license.

The remainder of the paper consists of sections where each of the
examples are explained in detail and a concluding section on features
of the library not covered in the examples, features that can be added
but have not been, and limitations of the numerical method. We end
this Introduction with a brief description of each of the examples.
In Section \ref{sec:Irreducible-flows} we compute the flows produced
by the leading terms of the spectral expansion of the active slip
for a single spherical particle away from boundaries. In Section \ref{sec:Effect-of-boundaries},
we examine the effect of boundaries - a plane wall and a plane liquid-liquid
interface - and show how the flows in the first example are altered.
In Section \ref{sec:Brownian-dynamics-with} we simulate the Brownian
motion of a pair of hydrodynamically interacting active particles
in a thermally fluctuating fluid confined by a plane wall. We identify
an attractive drag force from the active flow and an unbinding transition
with increasing temperature as entropic repulsion overwhelms this
active hydrodynamic attraction. In Section \ref{sec:Flow-induced-phase-separation}
we simulate the flow-induced phase separation (FIPS) of active particles
at a plane wall. This provides a quantitative description of the crystallization
of active particles that swim into a wall \citep{theurkauff2012dynamic,palacci2013living,buttinoni2013DynamicClustering,chen2015dynamic,petroff2015fast,thutupalli2018FIPS,aubret2018targeted}.
In Section \ref{sec:Irreducible-flows} we introduce a phoretic flux
on the particle surface and compute the phoretic field that it produces,
both away from and in the proximity of a no-flux wall. In Section
\ref{sec:Phoretic-arrest-of}, we show that a competition between
the hydrodynamic and phoretic interactions of autophoretic particles
can arrest the phase separation induced by flow alone \citep{singh2019competing}.
In the penultimate section, we show that the cost of computation increases
quadratically with the number of particles and decreases linearly
with the number of computational threads. 

\section{Mathematical underpinnings}

Our method relies on the reduction of linear elliptic partial differential
equations (PDE) to systems of linear algebraic equations. The four
key mathematical steps underpinning it are illustrated in this diagram:
\[
\boxed{\text{elliptic PDE}\xrightarrow{\,\,1\,\,}\text{boundary integral}\xrightarrow{\,\,2\,\,}\text{spectral expansion}\xrightarrow{\,\,3\,\,}\text{Ritz-Galerkin discretization}\xrightarrow{\,\,4\,\,}\text{truncation}}
\]
The \emph{first} step is the representation of the solution of an
elliptic PDE in a three-dimensional volume $V$ as an integral over
the boundary of the volume $S$ \citep{fkg1930bandwertaufgaben,ladyzhenskaya1969,youngren1975stokes,zick1982stokes,pozrikidis1992,muldowney1995spectral,cheng2005heritage,singh2015many}.
For the Laplace equation, this is the classical theorem of Green \citep{jackson1962classical};
for the Stokes equation, it is the generalization obtained by Lorentz
\citep{fkg1930bandwertaufgaben,lorentz1896eene,ladyzhenskaya1969}.
The integral representation leads to a linear integral equation that
provides a functional relation between the field and its flux on $S$.
Thus, if the surface flux in the Laplace equation is specified, the
surface concentration is determined by the solution of the Laplace
boundary integral equation. Similarly, if the surface velocity in
the Stokes equation is specified, the surface traction is determined
by the solution of the Stokes boundary integral equation. This transformation
of the governing PDEs is the most direct way of relating boundary
conditions (surface flux, slip velocities) to boundary values (surface
concentration, surface traction). It reduces the dimensionality of
the problem from a three-dimensional one in $V$ to a two-dimensional
one on $S$. The \emph{second} step is the spectral expansion of the
field and its flux in terms of global basis functions on $S$. We
use the geometry-adapted tensorial spherical harmonics, which provide
an unified way of expanding both scalar and vector quantities on the
surface of a sphere. These functions are both complete and orthogonal
and provide representations of the three-dimensional rotation group
\citep{hess2015tensors}. Thus, symmetries of the active boundary
conditions can be represented in a straightforward and transparent
manner. The \emph{third} step is the discretization of the integral
equation using the procedure of Ritz and Galerkin \citep{boyd2001chebyshev,finlayson1966method},
which reduces it to an infinite-dimensional self-adjoint linear system
in the expansion coefficients. This exploits the orthogonality of
the basis functions on the sphere. The matrix elements of the linear
system can be evaluated analytically in terms of the Green's functions
of the respective elliptic equations. The \emph{fourth} step is the
truncation of the infinite-dimensional linear system to a finite-dimensional
one that can be solved by standard methods of linear algebra adapted
for self-adjoint systems \citep{saad2003iterative}. Analytical solution
can be obtained by Jacobi iteration, which is equivalent to Smoluchowski's
method of reflections. Numerical solutions can be obtained by the
conjugate gradient method, at a cost quadratic in the number of unknowns.
From this solution, we can reconstruct the field and the flux on the
boundary, use these to determine the fields in the bulk, and from
there, compute derived quantities. These steps have been elaborated
in several papers \citep{ghose2014irreducible,singh2015many,singh2016crystallization,singh2018generalized,singh2019competing}
and we do not repeat them in detail here. Instead, we show below how
the method is applied to problems of experimental interest. 

\section{Library structure}

The overall organization of the library is show in Table \ref{tab:PyLaplaceStokes}
and will be referred to throughout the remainder of the paper. The PyStokes
library solves, respectively, the Stokes and Laplace equations, using
the reduction method explained in the previous section. The library
takes as input a set of expansion coefficients $\mathbf{J}_{i}^{(l)}$
for the prescribed active flux $j^{A}$ on the surface of the $i$-th
particle and computes the expansion coefficients of the surface concentration.
The active slip velocity $\boldsymbol{v}^{\mathcal{A}}$ is obtained
from this using the linear coupling relation $\boldsymbol{v}^{\mathcal{A}}=\mu_{c}\boldsymbol{\nabla}_{s}c$
\citep{anderson1989colloid}. The library outputs the expansion coefficients
$\mathbf{V}_{i}^{(l\sigma)}$ of the active slip. The PyStokes library
takes as input these expansion coefficients, which may also be specified
independently, and any body forces $\mathbf{F}_{i}^{B}$ and body
torques $\mathbf{T}_{i}^{B}$ acting on the particles and returns
their rigid body motion in terms of the velocities $\mathbf{V}_{i}$
and angular velocities $\boldsymbol{\Omega}_{i}$. In addition to
the joint computation of phoretic and hydrodynamic interactions, the
PyStokes library can be used to compute the hydrodynamically interacting
motion of squirming particles where the slip is specified independently
of a phoretic field, or the dynamics of passive sus- pensions where
the slip vanishes and forces and torques are prescribed. The PyStokes
library can also compute hydrodynamically correlated Brownian motion,
and thus, allows the study of the interplay between passive, active,
and Brownian contributions to motion. The library optionally computes
the corresponding field in $V$, necessary for insight and visualization.
Additionally, PyStokes computes the dissipation of mechanical energy
and the rheological response of the suspension. 
\begin{table*}[t]
\global\long\def\arraystretch{4}%
\includegraphics[width=0.75\textwidth]{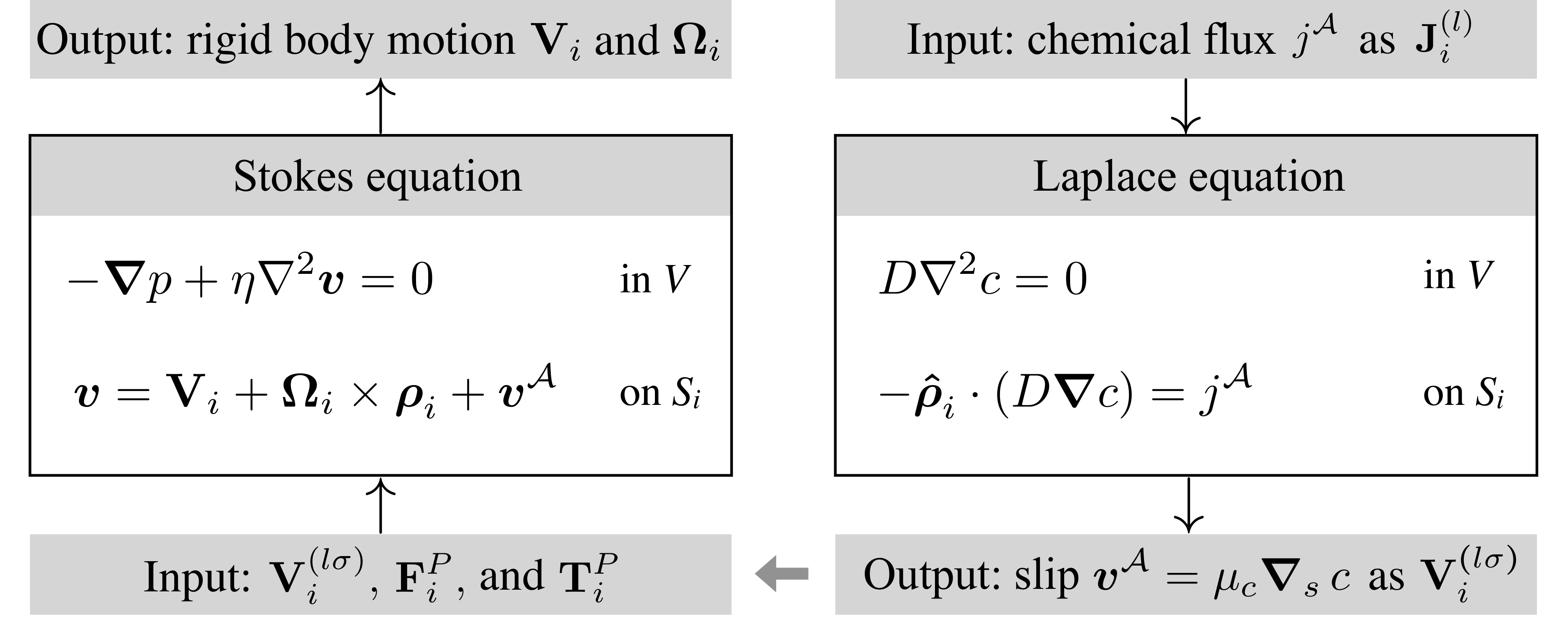}

\caption{\label{tab:PyLaplaceStokes}This schematic shows the governing equations
that determine the hydrodynamic and phoretic interactions between
active particles in a three-dimensional domain $V$. The equations
are coupled by the active boundary conditions on the surface $S_{i}$ of the $i$-th particle. The library takes
as input surface fluxes $j^{\mathcal{A}}$, specified in terms of
coefficients $\mathbf{J}_{i}^{(l)}$ of its tensorial harmonic expansion,
and returns slip velocities $\boldsymbol{v}^{\mathcal{A}}$ specified
in terms of coefficients $\mathbf{V}_{i}^{(l\sigma)}$ of its tensorial
harmonic expansion. To solve the Stokes part of the problem, the library
takes input these slip velocities and possible body forces $\mathbf{F}_{i}^{B}$
and torques $\mathbf{T}_{i}^{B}$ and returns the velocities $\mathbf{V}_{i}$
and angular velocities $\boldsymbol{\Omega}_{i}$. With slip velocities
set to zero, PyStokes computes the hydrodynamically interacting motion
of passive particles. Each library additionally computes the corresponding
fields in the bulk and quantities derived from these, like the entropy
production and rheological response. Particle indices are $i=1,\ldots,N$
and harmonic indices are $l=1,2,\ldots$ and $\sigma=s,a,t$ (see
text). }
\end{table*}

\section{Example 1 - Irreducible Active flows \label{sec:Irreducible-flows}}

Our first example shows how to plot the irreducible parts of an active
flow around a spherical particle, which we take to be far removed
from boundaries. For this example, we will take the reader step by
step from the governing PDE to the expressions for the irreducible
flows that the PyStokes library evaluates and plots. 
\begin{enumerate}
\item \emph{Elliptic PDE}\textbf{\emph{.}} The flow field $\boldsymbol{v}(\boldsymbol{r})$
satisfies the Stokes equation in the region $V$ exterior to the sphere
(radius $b$, centered at $\boldsymbol{R}$, oriented along unit vector
$\boldsymbol{p}$). On the sphere surface $S,$ it satisfies the slip
boundary condition
\begin{equation}
\lim_{\boldsymbol{r}\rightarrow S}\boldsymbol{v}(\boldsymbol{r})=\mathbf{V}+\boldsymbol{\Omega}\times\boldsymbol{\rho}+\boldsymbol{v}^{\mathcal{A}}(\boldsymbol{\rho)},
\end{equation}
where $\mathbf{V}$ and $\text{\ensuremath{\boldsymbol{\Omega}} are the rotational and translational velocities of the sphere, \ensuremath{\boldsymbol{v}^{\mathcal{A}}(\boldsymbol{\rho})} }$is
the slip velocity and and $\boldsymbol{\rho}$ is the radius vector
from the center to $S$. The fundamental solution of the Stokes equation
is given by the system\begin{subequations}
\begin{gather}
-\nabla_{\alpha}P_{\beta}+\eta\nabla^{2}G_{\alpha\beta}=-\delta\left(\boldsymbol{r}-\boldsymbol{r'}\right)\delta_{\alpha\beta},\\
K_{\alpha\beta\gamma}=-\delta_{\alpha\gamma}P_{\beta}+\eta\left(\nabla_{\gamma}G_{\alpha\beta}+\nabla_{\alpha}G_{\beta\gamma}\right),\quad\nabla_{\alpha}G_{\alpha\beta}=0.
\end{gather}
\end{subequations}where $G_{\alpha\beta}$ is the Green's function,
$P_{\alpha}$ is the pressure vector, $K_{\alpha\beta\gamma}$ is
the fundamental solution for the stress tensor, and $\eta$ is the
fluid viscosity. The Green's function may need to satisfy additional
boundary conditions which we keep unspecified for now. 
\item \emph{Boundary integral. }The fundamental solution, together with
the Lorentz reciprocal relation gives the boundary integral representation
\begin{alignat}{1}
v_{\alpha}(\boldsymbol{r})=-\int G_{\alpha\beta}(\boldsymbol{r},\boldsymbol{R}+\boldsymbol{\rho})\,f_{\beta}(\boldsymbol{\rho})\,d\text{S} & +\int K_{\beta\alpha\gamma}(\boldsymbol{r},\boldsymbol{R}+\boldsymbol{\rho})\hat{\rho}_{\gamma}v_{\beta}^{\mathcal{}}(\boldsymbol{\rho})\,d\text{S},\label{eq:BIE-v}
\end{alignat}
which expresses the flow field in $V$ in terms of a ``single-layer''
integral involving the traction and a ``double-layer'' integral
involving the boundary velocity. The latter is specified by the boundary
condition; the former must be determined in terms of it.
\item \emph{Spectral expansion. }The analytical evaluation of the two integrals
is possible if the slip and the traction are expanded spectrally in
terms of tensorial spherical harmonics,
\begin{gather}
\boldsymbol{v}^{\mathcal{A}}(\boldsymbol{\rho})=\sum_{l=1}^{\infty}w_{l-1}\mathbf{V}^{(l)}\cdot\mathbf{Y}^{(l-1)}(\bm{\hat{\rho}}),\quad\boldsymbol{f}(\boldsymbol{\rho})=\sum_{l=1}^{\infty}\tilde{w}_{l-1}\mathbf{F}^{(l)}\cdot\mathbf{Y}^{(l-1)}(\bm{\hat{\rho}}),\label{eq:spectral}
\end{gather}
where ${\bf Y}^{(l)}(\boldsymbol{\hat{\rho}})=(-1)^{l}\rho^{l+1}\nabla^{l}\rho^{-1}$
is the $l$-th irreducible tensorial harmonic. The $l$-th rank tensorial
coefficients $\mathbf{V}^{(l)}$ and $\mathbf{F}^{(l)}$ are symmetric
and irreducible in their last $l-1$ indices and have the dimensions
of velocity and force respectively. The dot indicates a complete contraction
of the indices of \textrm{${\bf Y}^{(l)}$ with the contractible indices
of the coefficients. }The $l$-dependent expansion weights are $w_{l}=\frac{1}{l!(2l-1)!!}$
and $\tilde{w}_{l}=\frac{2l+1}{4\pi b^{2}}$. The orthogonality of
the tensorial harmonics implies that\textrm{
\begin{alignat}{1}
\mathbf{V}^{(l)}=\tilde{w}_{l-1}\int\boldsymbol{v}^{\mathcal{A}}(\boldsymbol{R}+\bm{\rho})\mathbf{Y}^{(l-1)}(\bm{\hat{\rho}})d\text{S},\qquad\mathbf{F}^{(l)} & =w_{l-1}\int\boldsymbol{f}(\boldsymbol{R}+\bm{\rho})\mathbf{Y}^{(l-1)}(\bm{\hat{\rho}})d\text{S}.
\end{alignat}
The integral of the traction, $\mathbf{F}^{(1)}$, is the net hydrodynamic
force and the integral of the cross-product of the traction with the
radius vector, the antisymmetric part of ${\bf F}^{(2)},$ is the
net hydrodynamic torque \citep{singh2018generalized}. }
\item \emph{Ritz-Galerkin discretization. }Letting the point $\boldsymbol{r}$
approach $S$ from $V$ and matching the flow in the integral representation
with the prescribed boundary condition leads to an integral equation
for the traction. Multiplying both sides of the integral equation
by the $l$-th harmonic and integrating yields an infinite-dimensional
system of linear equations for the coefficients of the traction,
\begin{alignat}{1}
 & \tfrac{1}{2}\tilde{\mathbf{V}}^{(l)}=-\boldsymbol{G}^{(l,l')}\cdot\mathbf{F}^{(l')}+\boldsymbol{K}^{(l,l')}\cdot\tilde{\mathbf{V}}^{(l)},\label{eq:linear-system}
\end{alignat}
where repeated indices are summed over, $\tilde{\mathbf{V}}^{(1)}=\mathbf{V}+\mathbf{V}^{(1)},$
$\tilde{\mathbf{V}}^{(2)}=b\mathbf{\boldsymbol{\epsilon}\cdot\boldsymbol{\Omega}}+\mathbf{V}^{(2)}$,
$\tilde{\mathbf{V}}^{(l)}=\mathbf{V}^{(l)}$ for $l>2$, and $\boldsymbol{G}^{(l,\,l')}$
and $\boldsymbol{K}^{(l,\,l')}$ are the matrix elements of the linear
system given in terms of integrals of the Green's function and the
fundamental stress solution. These can be evaluated analytically \citep{singh2018generalized}.
\item \emph{Truncation: }The infinite-dimensional linear system has to be
truncated to a finite-dimensional one for tractability. We truncate
the system at $l=2$ and decompose the coefficients into their irreducible
symmetric ($s)$, antisymmetric $(a)$ and trace $(t)$ parts, so
that slip and the traction are\begin{subequations}
\begin{eqnarray}
\boldsymbol{v}^{\mathcal{A}}(\bm{\rho})=\mathbf{V}^{(1s)}+[\mathbf{V}^{(2s)}-\boldsymbol{\epsilon}\cdot\mathbf{V}^{(2a)}]\cdot\mathbf{Y}^{(1)}+\tfrac{1}{6}[\mathbf{V}^{(3s)}-\tfrac{2}{3}\boldsymbol{\Delta}\cdot(\boldsymbol{\boldsymbol{\epsilon}}\cdot\mathbf{V}^{(3a)})+\tfrac{3}{5}\boldsymbol{\Delta}\cdot(\boldsymbol{\delta}\mathbf{V}^{(3t)})]\cdot\mathbf{Y}^{(2)},\label{eq:slip-truncation-1}\\
4\pi b^{2}\,\boldsymbol{f}(\bm{\rho})=\mathbf{F}^{(1s)}+3[\mathbf{F}^{(2s)}-\tfrac{1}{2}\boldsymbol{\epsilon}\cdot\mathbf{F}^{(2a)}]\cdot\mathbf{Y}^{(1)}+5[\mathbf{F}^{(3s)}-\tfrac{2}{3}\boldsymbol{\Delta}\cdot(\boldsymbol{\boldsymbol{\epsilon}}\cdot\mathbf{F}^{(3a)})+\tfrac{3}{5}\boldsymbol{\Delta}\cdot(\boldsymbol{\delta}\mathbf{F}^{(3t)})]\cdot\mathbf{Y}^{(2)}.
\end{eqnarray}
\end{subequations}Here $\boldsymbol{\epsilon}$ and $\boldsymbol{\delta}$
are the Levi-Civita and Kronecker tensors and $\Delta_{\alpha\beta\mu\nu}=\tfrac{1}{2}(\delta_{\alpha\nu}\delta_{\beta\mu}+\delta_{\alpha\mu}\delta_{\beta\nu}-\frac{2}{3}\delta_{\alpha\beta}\delta_{\mu\nu})$
symmetrises and detraces second-rank tensors. This truncation includes
all long-ranged contributions to the active flow and is sufficient
to parametrize experimentally measured active flows around microorganisms,
active drops, and autophoretic colloids \citep{ghose2014irreducible,singh2016crystallization,thutupalli2018FIPS}.
The solution of the finite-dimensional linear system yields a linear
relation between the irreducible force and velocity coefficients,
the ``generalized Stokes laws'', which are $\mathbf{F}^{(l\sigma)}=-\boldsymbol{\gamma}^{(l\sigma,l'\sigma')}\cdot\tilde{\mathbf{V}}^{(l'\sigma')}$
where $\sigma=s,a,t$ and repeated indices are summed over. In an
unbounded domain the friction tensors $\boldsymbol{\gamma}^{(l\sigma,l'\sigma')}$
take on a particularly simple form: they are diagonal in both the
$l$ and $\sigma$ indices, $\boldsymbol{\gamma}^{(l\sigma,l'\sigma')}\equiv\delta_{ll'}\delta_{\sigma\sigma'}\gamma^{l\sigma}\Delta^{(l)}$
so that a single scalar $\gamma^{l\sigma}$ determines them. For $l\sigma=1s$
and $l\sigma=2a$ these are the familiar coefficients $6\pi\eta b$
and $8\pi\eta b^{3}$ that appear in Stokes laws for the force and
torque. 
\end{enumerate}
\begin{figure}[H]
\begin{lstlisting}[language=Python,basicstyle={\scriptsize\ttfamily},breaklines=true]
# ex1.py - flow around an active colloid in an unbounded domain
import pystokes, numpy as np, matplotlib.pyplot as plt 

# particle radius, fluid viscosity, and number of particles 
b, eta, Np = 1.0, 1.0/6.0, 1

#initialise position, orientation and body force on the colloid 
r, p, F = np.array([0.0, 0.0, 0.0]), np.array([0.0, -1.0, 0]),  np.array([0.0, 1.0, 0])

# irreducible coeffcients 
V2s = pystokes.utils.irreducibleTensors(2, p) 
V3t = pystokes.utils.irreducibleTensors(1, p) 

# space dimension , extent , discretization 
dim, L, Ng = 3, 10, 100;

# instantiate the Flow class 
flow = pystokes.unbounded.Flow(radius=b, particles=Np, viscosity=eta, gridpoints=Ng*Ng) 


# plot using subplots on a grid
rr, vv = pystokes.utils.gridXY(dim, L, Ng)

plt.figure(figsize=(15, 10), edgecolor='gray', linewidth=4)
plt.subplot(231);  vv=vv*0;  flow.flowField1s(vv, rr, r, F)   
pystokes.utils.plotStreamlinesXY(vv, rr, r, offset=6e-1, title='1s', density=2)

plt.subplot(232);   vv=vv*0;  flow.flowField2s(vv, rr, r, V2s)   
pystokes.utils.plotStreamlinesXY(vv, rr, r, offset=4e-2, title='2s', density=2)

plt.subplot(233); vv=vv*0;  flow.flowField3t(vv, rr, r, V3t)   
pystokes.utils.plotStreamlinesXY(vv, rr, r, offset=4e-4, title='3t', density=2)
\end{lstlisting}
\includegraphics[width=1\textwidth]{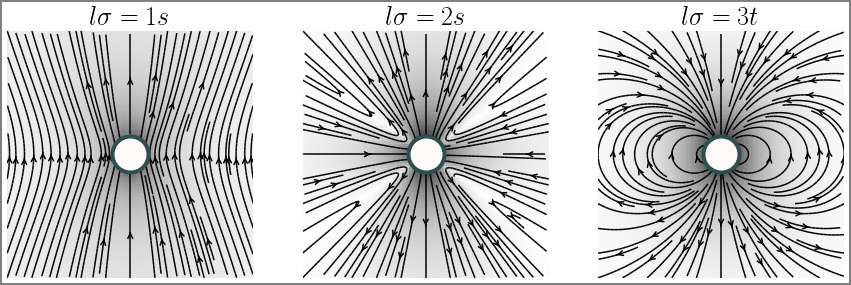}\caption{Irreducible flows: Streamlines of the fluid overlaid on the pseudocolor
plot of the logarithm of the flow speed (increasing in strength from
light to dark). The first panel is the flow due the $1s$ mode of
the traction, the second and third panels are due to $2s$ and $3t$
modes of the active slip. The streamlines inherit the symmetry of
the modes. \label{fig:1}}
\end{figure}
Inserting the truncated spectral expansions for the slip and traction
in the boundary integral, eliminating the unknown traction coefficients
in favour of the known slip coefficients, expanding the Green's function
about the center of the sphere and finally using the orthogonality
of the tensorial harmonics, we obtain the flow due to each irreducible
slip mode as\begin{subequations}
\begin{gather}
\text{}\boldsymbol{v}^{1s}(\boldsymbol{r})=-(1+\tfrac{b^{2}}{6}\nabla^{2})\,\mathbf{G}\cdot\mathbf{F}^{(1s)},\label{eq:irredFlow}\\
\boldsymbol{v}^{2s}(\boldsymbol{r})=\tfrac{28\pi\eta b^{2}}{3}(1+\tfrac{b^{2}}{10}\nabla^{2})\,\boldsymbol{\nabla}\mathbf{G}\cdot\mathbf{V}^{(2s)},\qquad\boldsymbol{v}^{2a}(\boldsymbol{r})=-\frac{1}{2}(\boldsymbol{\nabla}\times\mathbf{G})\cdot\mathbf{F}^{(2a)},\\
\boldsymbol{v}^{3s}(\boldsymbol{r})=\tfrac{13\pi\eta b^{3}}{9}(1+\tfrac{b^{2}}{14}\nabla^{2})\boldsymbol{\nabla}\boldsymbol{\nabla}\mathbf{G}\cdot\mathbf{V}^{(3s)},\quad\boldsymbol{v}^{3a}(\boldsymbol{r})=\tfrac{2\pi\eta b^{3}}{3}\boldsymbol{\nabla}(\boldsymbol{\nabla}\times\mathbf{G})\cdot\mathbf{V}^{(3a)},\quad\boldsymbol{v}^{3t}(\boldsymbol{r})=\tfrac{2\pi\eta b^{3}}{5}\nabla^{2}\mathbf{G}\cdot\mathbf{V}^{(3t)}.
\end{gather}
\end{subequations}We emphasise that these expressions are valid for
any Green's function of the Stokes equation, provided they satisfy
the additional boundary conditions that may be imposed. We also note
that no discretization of space, either in $V$ or on $S$ is involved,
making the result ``grid-free''. The $\mathtt{flow}$ class in PyStokes
computes these expressions for supplied values of the force, the torque
and slip coefficients. For this example, we choose an unbounded domain
with the flow vanishing at infinity, for which the Green's function
is the Oseen tensor,

\begin{equation}
G_{\alpha\beta}^{\text{o}}(\boldsymbol{r}-\boldsymbol{r}^{\prime})=\frac{1}{8\pi\eta}\left(\nabla^{2}\delta_{\alpha\beta}-\nabla_{\alpha}\nabla_{\beta}\right)|\boldsymbol{r}-\boldsymbol{r}^{\prime}|.\label{eq:OseenTensor}
\end{equation}
The code listed in Fig.(\ref{fig:1}) computes the irreducible flows
$1s$, $2s,$ and $3t$ for radius $b=1$, viscosity $\eta=1/6$,
location $\boldsymbol{R}=(0,0,0)$ and orientation $\boldsymbol{p}=(0,-1,0).$
The coefficients are parametrised as $F_{\alpha}^{(1s)}=-p_{\alpha},$
$V_{\alpha\beta}^{(2s)}=p_{\alpha}p_{\beta}-\frac{1}{3}\delta_{\alpha\beta}$
and $V_{\alpha}^{(3t)}=p_{\alpha}$. This information is supplied
to the $\mathtt{flow}$ class which is instantiated for an unbounded
fluid. The $l\sigma$ irreducible component of the flow is computed
by the calling the function $\mathtt{flow.flowfieldl\sigma}$. This
is passed to a generic plotting function to compute the streamlines
in a plane of symmetry. These are shown in the code output where the
polar symmetries of the $1s$ and $3t$ modes and the nematic symmetry
of the $2s$ mode can be seen clearly. For vanishing radius, these
are the Stokeslet, potential dipole and stresslet singularities \citep{batchelor2000introduction}.
\begin{figure}[H]
\begin{lstlisting}[language=Python,basicstyle={\scriptsize\ttfamily},breaklines=true]
# ex2.py - comparison of flow around an active colloid near a plane surface (wall and interface)
import pystokes, numpy as np, matplotlib.pyplot as plt 

# particle radius, fluid viscosity, and number of particles 
b, eta, Np = 1.0, 1.0/6.0, 1

#initialise position, orientation and body force on the colloid 
r, p, F = np.array([0.0, 0.0, 3.4]), np.array([0.0, 0.0, 1]),  np.array([0.0, 0.0, 1])

# irreducible coeffcients 
V2s = pystokes.utils.irreducibleTensors(2, p) 
V3t = pystokes.utils.irreducibleTensors(1, p)

# space dimension , extent , discretization 
dim, L, Ng = 3, 10, 64; 

# Instantiate the Flow class 
wFlow = pystokes.wallBounded.Flow(radius=b, particles=Np, viscosity=eta, gridpoints=Ng*Ng)
iFlow = pystokes.interface.Flow(radius=b, particles=Np, viscosity=eta, gridpoints=Ng*Ng)

# plot using subplots on a given grid
plt.figure(figsize=(15, 8))
rr, vv = pystokes.utils.gridYZ(dim, L, Ng)

plt.subplot(231);    vv=vv*0;    wFlow.flowField1s(vv, rr, r, F)   
pystokes.utils.plotStreamlinesYZsurf(vv, rr, r, mask=0.5, title='1s')
plt.subplot(232);    vv=vv*0;    wFlow.flowField2s(vv, rr, r, V2s)   
pystokes.utils.plotStreamlinesYZsurf(vv, rr, r, mask=0.4, title='2s')
plt.subplot(233);    vv=vv*0;    wFlow.flowField3t(vv, rr, r, V3t)   
pystokes.utils.plotStreamlinesYZsurf(vv, rr, r, mask=0.5, title='3t')
plt.subplot(234);    vv=vv*0;    iFlow.flowField1s(vv, rr, r, F)   
pystokes.utils.plotStreamlinesYZsurf(vv, rr, r, mask=0.0, title='None')
plt.subplot(235);    vv=vv*0;    iFlow.flowField2s(vv, rr, r, V2s)   
pystokes.utils.plotStreamlinesYZsurf(vv, rr, r, mask=0.0, title='None')
plt.subplot(236);    vv=vv*0;    iFlow.flowField3t(vv, rr, r, V3t)   
pystokes.utils.plotStreamlinesYZsurf(vv, rr, r, mask=0.5, title='None')
\end{lstlisting}
\includegraphics[width=1\textwidth]{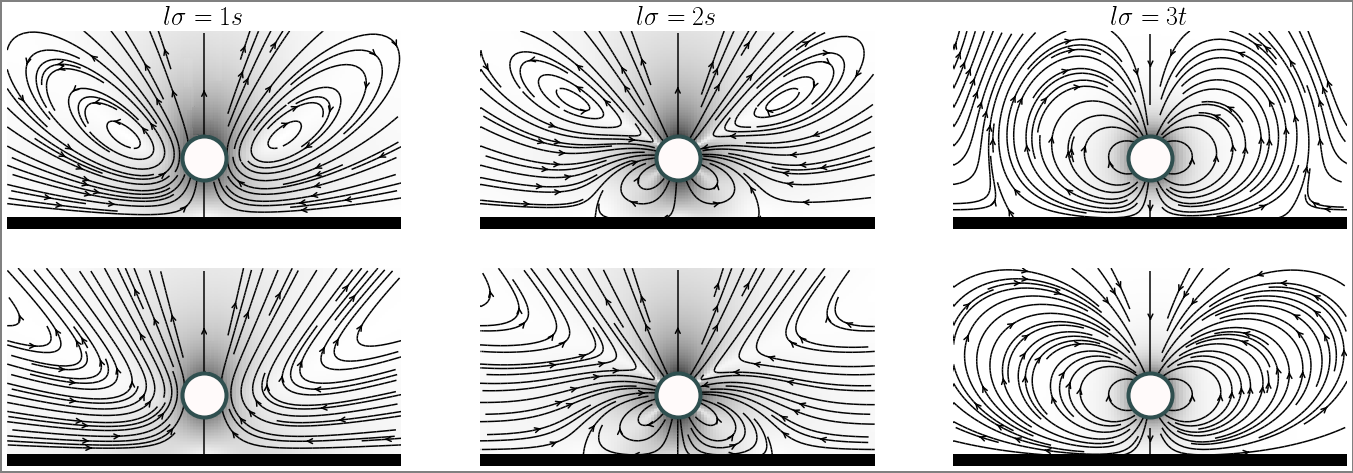}\caption{Distortion of irreducible flows, by a plane no-slip wall (top row)
and plane no-shear interface (bottom row), due to the $1s$ mode of
the traction and $2s$ and $3t$ modes of the active slip. The streamlines
no longer inherit the symmetry of the modes. \label{fig:2}}
\end{figure}

\section{Example 2 - Effect of plane boundaries \label{sec:Effect-of-boundaries}}

Our second example illustrates how irreducible flows are modified
by the proximity to plane boundaries. This is of relevance to experiments,
where confinement by boundaries is commonplace \citep{goldstein2015green,thutupalli2018FIPS}.
This also illustrates the flexibility of our method, as the only quantity
that needs to be changed is the Green's function. The Green's function
for a no-slip wall is the Lorentz-Blake tensor
\begin{eqnarray}
G_{\alpha\beta}^{\text{w}}(\boldsymbol{R}_{i},\,\boldsymbol{R}_{j}) & = & G_{\alpha\beta}^{\text{o}}(\boldsymbol{r}_{ij})-G_{\alpha\beta}^{\text{o}}(\boldsymbol{r}_{ij}^{*})-2h\nabla_{{\scriptscriptstyle \boldsymbol{r}_{\gamma}^{*}}}G_{\alpha3}^{\text{o}}(\boldsymbol{r}_{ij}^{*})\mathcal{M}_{\beta\gamma}+h^{2}\nabla_{{\scriptscriptstyle \boldsymbol{r}^{*}}}^{2}G_{\alpha\gamma}^{\text{o}}(\boldsymbol{r}_{ij}^{*})\mathcal{M}_{\beta\gamma}.\label{eq:stokesGwall}
\end{eqnarray}
Here\textrm{ }$\boldsymbol{r}_{ij}^{*}=\mathbf{\boldsymbol{R}}_{i}-\mathbf{\boldsymbol{R}}_{j}^{*}$,
where \textrm{$\boldsymbol{R}_{j}^{*}=\boldsymbol{\mathcal{M}}\cdot\boldsymbol{R}$
is the image of the $j$-th colloid at a distance $h$ from plane
boundary} and\textrm{ $\boldsymbol{\mathcal{M}}=\boldsymbol{I}-2\mathbf{\hat{z}}\mathbf{\hat{z}}$}
is the reflection operator. The Green's function for a no-shear plane
air-water interface is
\begin{eqnarray}
G_{\alpha\beta}^{\text{i}}(\boldsymbol{R}_{i},\,\boldsymbol{R}_{j}) & = & G_{\alpha\beta}^{\text{o}}(\boldsymbol{r}_{ij})+(\delta_{\beta\rho}\delta_{\rho\gamma}-\delta_{\beta3}\delta_{3\gamma})G_{\alpha\gamma}^{\text{o}}(\boldsymbol{r}_{ij}^{*}).
\end{eqnarray}
The plane boundary is placed at $z=0$ and the flows are plotted in
the half-space $z>0$. The irreducible flows for each boundary condition
are obtained by evaluated Eq.(8) with the corresponding Green's function.
The irreducible flows for the modes in Example 1 are shown in Fig.(\ref{fig:2}),
with no-slip wall in the top panels and no-shear interface in the
bottom panels. We use the same initialization as in Example 1, but
instantiate wall-bounded and interfacially-bounded classes $\mathtt{wflow}$
and $\mathtt{iflow}$ respectively. Though arbitrary viscosity ratios
are allowed, PyStokes assumes an air-water interface as default, setting
the viscosity ratio between the two fluids to zero, as in this example.
The $l\sigma$ irreducible component of the flow is computed by the
calling the function $\mathtt{wflow.flowfieldl\sigma}$ and passed
to a generic plotting function for streamline computation. Note that
the streamlines near an interface do not close, unlike those near
a plane wall. The perpendicular component of the flow near a wall
is about an order of magnitude larger than that near an interface.
These features have been recently used to understand the control by
boundaries of flow-induced phase separation of active particles \citep{thutupalli2018FIPS}. 

\section{Example 3 - Active Brownian Hydrodynamics\label{sec:Brownian-dynamics-with}}

\begin{figure}
\begin{lstlisting}[language=Python,basicstyle={\scriptsize\ttfamily},breaklines=true]
# ex3.py: competition between flow-induced forces and thermal fluctuations
import pystokes, numpy as np, matplotlib.pyplot as plt

# particle radius, self-propulsion speed, number and fluid viscosity 
b, vs, Np, eta = 1.0, 0.5, 2, 0.1

rbm     = pystokes.wallBounded.Rbm(radius=b, particles=Np, viscosity=eta)
forces  = pystokes.forceFields.Forces(particles=Np)

def twoBodyDynamics(T=1):     
	"""simulation of two active colloid near a wall in a fluid at temperature T"""          
	#initial position and orientation     
	r, p = np.array([-2.5,2.5, 0,0 , 2.5, 2.5]), np.array([0,0, 0,0, -1.0,-1.0])          
	
	# integration parameters and arrays     
	Nt=2**17;  x1=np.zeros(Nt);  x2=np.zeros(Nt)     
	x1[0], x2[0] = r[0], r[1];  dt=0.01;  sqdt=np.sqrt(T*dt)     
	F = np.zeros(3*Np);  v = np.zeros(3*Np); vv = np.zeros(3*Np)     
	F0 = 6*np.pi*eta*b*vs*(1+9*b/(8*r[4])); #active stall force
	
	# integration loop     
	for i in range(Nt-1):         
	forces.lennardJones(F,r,lje=.12,ljr=2.5); F[4],F[5]= F0, F0         
	rbm.mobilityTT(v, r, F);  rbm.calcNoiseMuTT(vv, r)                                 
	
	# Euler-Maryuama integration         
	x1[i+1] = x1[i] + dt*v[0] + sqdt*vv[0]         
	x2[i+1] = x2[i] + dt*v[1] + sqdt*vv[1]
        
	#reset the variables for next time step         
	r[0],r[1],v,vv,F[0:3] = x1[i+1], x2[i+1],v*0,vv*0,F[0:3]*0             
	return x1, x2

# dynamics as a function of temperature
T=([0, .1]);   pystokes.utils.plotTrajectory(twoBodyDynamics, T)
\end{lstlisting}
\includegraphics[width=1\textwidth]{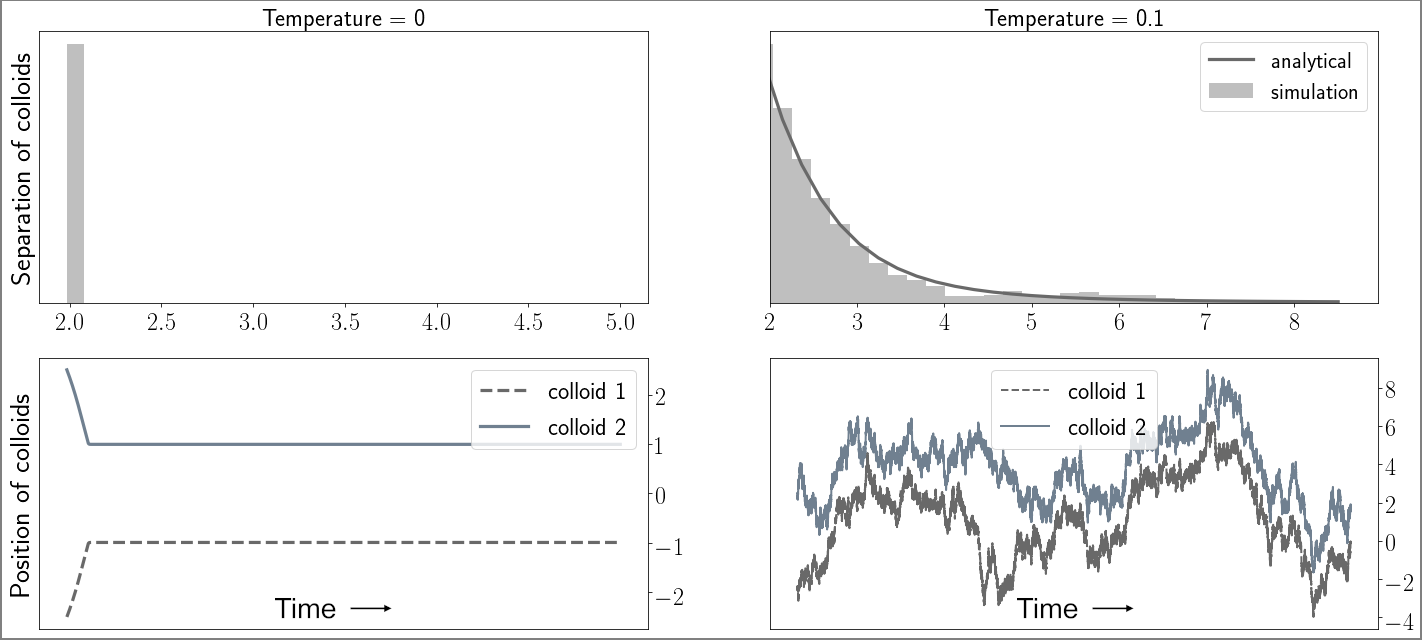}\caption{Fluctuating hydrodynamics of a pair of active particles near a plane
wall. The two particles form a bound state in absence of thermal fluctuations.
At finite temperatures, they ``unbind'' and distribution of their
separation follows from an nonequilibrium potential (see text).\label{fig:3}}
\end{figure}
Our third example shows how to simulate the dynamics of active particles
including active, passive and thermal forces. This example assumes
that orientational degrees of freedom are not dynamical, as is often
the case with strongly bottom-heavy active particles. The PyStokes
library computes the rigid body motion of the active particles consistent
with the overdamped Langevin equation
\begin{equation}
\mathbf{F}_{i}^{H}+\mathbf{F}_{i}^{B}+\hat{\mathbf{F}}_{i}=0,
\end{equation}
where $\mathbf{F}_{i}^{H}=\mathbf{F}_{i}^{(1s)}$, $\mathbf{F}_{i}^{B}$
and $\hat{\mathbf{F}}_{i}$ are the hydrodynamic, body and thermal
forces respectively and $i$ is the particle index. We now consider
a minimal model of the slip, retaining only the lowest modes of vectorial
symmetry:
\begin{equation}
\boldsymbol{v}^{\mathcal{A}}(\boldsymbol{\hat{\rho}}_{i})=\mathbf{V}_{i}^{(1s)}+\tfrac{1}{15}\mathbf{V}_{i}^{(3t)}\cdot\mathbf{Y}^{(2)}(\boldsymbol{\hat{\rho}}_{i}).\label{eq:minimalSlip}
\end{equation}
We parametrize the coefficients of the slip uniaxially, in terms of
the orientation $\boldsymbol{p}_{i}$ of the $i$-th colloid and its
self-propulsion speed $v_{s}$, as
\begin{alignat}{1}
 & \mathbf{V}_{i}^{(1s)}=v_{s}\,\boldsymbol{p}_{i},\qquad\mathbf{V}_{i}^{(3t)}=V_{0}^{(3t)}\,\boldsymbol{p}_{i}.\label{eq:parametrization}
\end{alignat}
The hydrodynamic forces are then given by the generalized Stokes laws,
\begin{alignat}{1}
\mathbf{F}_{i}^{H}= & -\boldsymbol{\gamma}_{ij}^{TT}\cdot(\mathbf{V}_{j}+\mathbf{V}_{j}^{(1s)})-\boldsymbol{\gamma}_{ij}^{(T,\,3t)}\cdot\mathbf{V}_{j}^{(3t)}.\label{eq:linear-force-torque}
\end{alignat}
where sub-dominant rotational contributions have been neglected. The
explicit forms of the tensors $\boldsymbol{\gamma}_{ij}^{TT}$ and
$\boldsymbol{\gamma}_{ij}^{(T,l\sigma)}$ follows from the solution
of the many-particle version of the linear system given in Eq. (\ref{eq:linear-system}).
The off-diagonal terms, with $i\neq j$, represent hydrodynamic interactions
and are obtained as an infinite series in the Green's function and
its derivatives \citep{singh2018generalized}. With this, force balance
becomes
\begin{equation}
-\boldsymbol{\gamma}_{ij}^{TT}\cdot(\mathbf{V}_{j}+\mathbf{V}_{j}^{(1s)})-\boldsymbol{\gamma}_{ij}^{(T,\,3t)}\cdot\mathbf{V}_{j}^{(3t)}+\mathbf{F}_{i}^{B}+\hat{\mathbf{F}}_{i}=0.\label{eq:forceBalance}
\end{equation}
This shows that in the absence of slip modes with $l>1$, external
forces, thermal fluctuations and hydrodynamic interactions, the translational
velocity of the $i$-th particle is given by $-{\bf V}_{i}^{(1s)}$.
It is convenient to introduce the notation ${\bf V}_{i}^{\mathcal{A}}=-{\bf V}_{i}^{(1s)}$
and then solve the force balance equation for the translational velocity.
This gives the overdamped Langevin equation \citep{singh2017fluctuation,singh2018generalized}
\begin{alignat}{1}
\dot{\boldsymbol{R}}_{i} & =\underbrace{\boldsymbol{\mu}_{ij}^{TT}\cdot\mathbf{F}_{j}^{B}}_{\mathrm{Passive}}+\underbrace{\boldsymbol{\pi}_{ij}^{(T,\,l\sigma)}\cdot\mathbf{\mathsf{\mathbf{V}}}_{j}^{(3t)}+\mathsf{\mathbf{V}}_{i}^{\mathcal{A}}}_{\mathrm{Active}}+\underbrace{\sqrt{2k_{B}T\bm{\mu}_{ij}^{TT}}\cdot\bm{\xi}_{j}^{T}(t)}_{\text{Brownian}}\label{eq:RBM-angular-velocity}
\end{alignat}
which is the basis for our Brownian dynamics algorithm. Here $\bm{\xi}_{j}^{T}$
is a vector of zero-mean unit-variance independent Gaussian random
variables, the mobility matrix $\boldsymbol{\mu}_{ij}^{TT}$ is the
inverse of the friction tensor $\boldsymbol{\gamma}_{ij}^{TT}$ and
the propulsion tensor $\boldsymbol{\pi}_{ij}^{(T,3t)}=-\boldsymbol{\mu}_{ik}^{TT}\cdot\boldsymbol{\gamma}_{kj}^{(T,\,3t)}$. 

We now look at the dynamics of a pair of active particles, $i=1,2$,
near a no-slip wall. We assume a truncated Lennard-Jones two-body
interaction between the particles and a different truncated Lennard-Jones
one-body interaction with the wall, designed to prevent particle-particle
and particle-wall overlaps. The orientation is taken to point into
the wall and a sufficiently large torque is added to prevent re-orientation.
Vertical motion ceases when repulsion from the wall balances active
propulsion into it. Then, the net external force on the particle points
normally away from the wall. This implies that the leading contribution
to active flow is due to the ${\bf F}^{(1s)}$ mode, as shown in Fig.(\ref{fig:2}).
This flow drags neighboring particles into each other and leads to
the formation of a bound state \citep{singh2016crystallization}.
This is the basis for numerous aggregation phenomena of active particles
near walls and interfaces \citep{theurkauff2012dynamic,palacci2013living,buttinoni2013DynamicClustering,chen2015dynamic,petroff2015fast,thutupalli2018FIPS,aubret2018targeted}.
In terms of equations of motion, the $z-$component of force balance
now reads
\begin{gather}
-\gamma_{11}^{zz}v_{s}+F_{1z}^{B}+\xi_{1}^{z}=0,\qquad-\gamma_{22}^{zz}v_{s}+F_{2z}^{B}+\xi_{2}^{z}=0.
\end{gather}
where the first term is the active propulsive force in the $z$--direction,
the second term is the $z-$component of the net force, and the third
term is the noise. The solution of this equation implicitly gives
the mean height $\bar{h}$ above the wall at which the particles come
to rest. The $x-$component of force balance gives
\begin{gather}
-\gamma_{11}^{xx}\dot{x}_{1}-\gamma_{12}^{xx}\dot{x}_{2}+\gamma_{12}^{xz}v_{s}+F_{1x}^{B}+\xi_{1}^{x}=0,\qquad-\gamma_{21}^{xx}\dot{x}_{1}-\gamma_{22}^{xx}\dot{x}_{2}+\gamma_{21}^{xz}v_{s}+F_{2x}^{B}+\xi_{2}^{x}=0,
\end{gather}
where the friction tensors are now evaluated at the mean height and
the instantaneous separation between the particles. This effectively
decouples the vertical and horizontal components of motion. The first
two terms are the self- and mutual- Stokes drags, the third term is
the drag from the active hydrodynamic flow, the fourth term is the
body force and the last term is the noise. The above overdamped Langevin
equation can be written in standard Ito form as
\begin{align}
d\bfrac{x_{1}}{x_{2}} & =\bfrac{\mu_{11}^{xx}\quad\mu_{12}^{xx}}{\mu_{21}^{xx}\quad\mu_{22}^{xx}}\bfrac{\gamma_{12}^{xz}v_{s}+F_{1x}^{B}}{\gamma_{21}^{xz}v_{s}+F_{2x}^{B}}dt+\bfrac{\sigma_{11}\quad\sigma_{12}}{\sigma_{21}\quad\sigma_{22}}\bfrac{dW_{1}}{dW_{2}},\label{eq:ito}
\end{align}
where the matrix of variances containing $\sigma_{ij}$ is the Cholesky
factor of the matrix of mobilities $\mu_{ij}^{xx}$. These are the
equations we simulate in Example 3. The truncated Lennard-Jones interactions
are $U=\epsilon(\frac{r_{{\scriptscriptstyle {min}}}}{r})^{12}-2\epsilon(\frac{r_{{\scriptscriptstyle {min}}}}{r})^{6}+\epsilon,$
for separation $r<r_{{\scriptscriptstyle {min}}}$, and zero otherwise
\citep{weeks1971role}. In the absence of thermal fluctuations, the
pair form a bound state due to the attractive active drag but at sufficiently
large temperatures, they ``unbind'' due to entropic forces. Remarkably,
the distribution of their in-plane separation can be expressed in
Gibbsian form, with a non-equilibrium potential, as shown in Fig.
(\ref{fig:3}). We now explain why this is so. 
\begin{figure}[h]
\begin{lstlisting}[language=Python,basicstyle={\scriptsize\ttfamily},breaklines=true]
# ex4.py: flow-induced phase separation of active colloids at a wall
import pystokes, numpy as np, matplotlib.pyplot as plt

# particle radius, self-propulsion speed, number and fluid viscosity 
b, vs, Np, eta = 1.0, 1.0, 128, 1

# initialise 
r = pystokes.utils.initialConditionRandom(Np) # positions random in plane of wall
p = np.zeros(3*Np); p[2*Np:3*Np] = -1         # orientations pointing into wall

# instantiate
rbm   = pystokes.interface.Rbm(radius=b, particles=Np, viscosity=eta)
force = pystokes.forceFields.Forces(particles=Np) 

def rhs(rp):     
	"""right hand side of the rigid body motion equation (rbm)    
	rp: is the array of position and orientations of the colloids"""     
	# assign fresh values at each time step     
	r = rp[0:3*Np];   p = rp[3*Np:6*Np]     
	F, v, o = np.zeros(3*Np), np.zeros(3*Np), np.zeros(3*Np)          
	
	force.lennardJonesWall(F, r, lje=0.01, ljr=5, wlje=1.2, wljr=3.4)   
	rbm.mobilityTT(v, r, F)              
	
	V1s = vs*p;  V3t=0.6*V1s;     
	rbm.propulsionT3t(v, r, V3t);    v = v + V1s     
	return np.concatenate( (v,o) )

# simulate the resulting system 
Tf, Npts = 150, 200 
pystokes.utils.simulate(np.concatenate((r,p)), Tf,Npts,rhs,integrator='odeint', filename='crystallization')

# plot the data at specific time instants 
pystokes.utils.plotConfigs(t=[1, 40, 100, 200], ms=60, tau=(Tf/Npts)/(b/vs), filename='crystallization')
\end{lstlisting}
\includegraphics[width=1\textwidth]{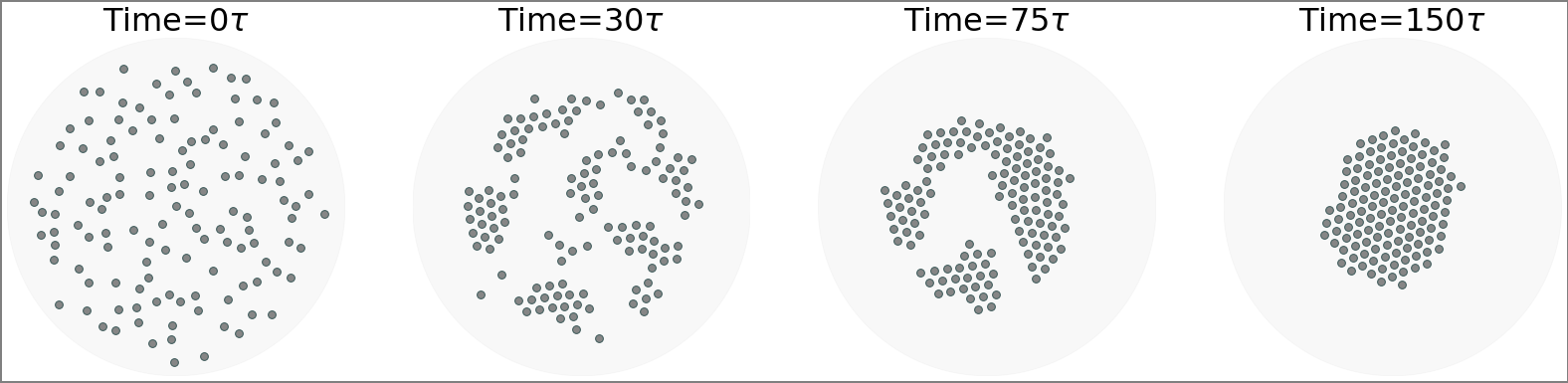}\caption{Flow-induced phase separation of active particles at a plane no-slip
wall. Starting from a non-crystalline distribution, active particles
crystallize into a single cluster due to long-ranged active hydrodynamic
interactions between them \citep{singh2016crystallization}.\label{fig:4}}
\end{figure}

Denoting the $x$-component of the active hydrodynamic drag as $F_{x}^{\mathcal{A}}$
and using the leading form for $\gamma_{21}^{xz}$ we have
\begin{equation}
F_{x}^{\mathcal{A}}=\gamma_{21}^{xz}v_{s}=\gamma_{{\scriptscriptstyle \parallel}}\gamma_{\perp}G_{xz}^{\text{w}}v_{s}=-\frac{\gamma_{{\scriptscriptstyle \parallel}}\gamma_{\perp}v_{s}}{2\pi\eta}\frac{3h^{3}}{(r^{2}+4\bar{h}^{2})^{5/2}}r^{x},
\end{equation}
where $\gamma_{{\scriptscriptstyle \parallel}}$ and $\gamma_{\perp}$
are the self-friction coefficients in the directions parallel ($\parallel$)
and perpendicular ($\perp$) to the wall \citep{kim2005}. This shows
that the active drag force can be written as the gradient of a potential
\begin{equation}
\Phi(\boldsymbol{r})=-\frac{\gamma_{{\scriptscriptstyle \parallel}}\gamma_{\perp}v_{s}}{2\pi\eta}\frac{h^{3}}{(r^{2}+4\bar{h}^{2})^{3/2}},
\end{equation}
whose strength depends on the propulsion speed. Potentials of identical
functional form, but with different prefactors, have been obtained
before for electrophoresis \citep{squires2001effective} and thermophoresis
\citep{di2009colloidal} without associating them to an active drag
force, as we have done here. In spite of the non-equilibrium origin
of the potential, it leads to a Gibbs distribution for the particle
positions, $P\sim\exp[-(\Phi+U)/k_{B}T]$, as the Ito equation satisfies
potential conditions whenever $\gamma_{12}^{xx}$, and by Onsager
symmetry $\gamma_{21}^{xx}$, is a gradient. This understanding of
the active Brownian hydrodynamics of a pair of bottom-heavy active
particles near a plane wall rationalizes the ubiquitously observed
crystallization of active particles near plane boundaries \citep{theurkauff2012dynamic,palacci2013living,buttinoni2013DynamicClustering,chen2015dynamic,petroff2015fast,thutupalli2018FIPS,aubret2018targeted},
which is our next example. 

\section{Example 4 - Flow-induced phase separation at a wall\label{sec:Flow-induced-phase-separation}}

Our fourth example demonstrates the flow-induced phase separation
(FIPS) of active particles at a plane no-slip wall. We use the same
model of active particles as described in the Example 3 but consider
a large number of them. As before, the slip is truncated to contain
only $1s$ and $3t$ modes of vectorial symmetry, which are parametrized
in term of the orientation $\boldsymbol{p}_{i}$ of the colloids,
see Eqs.(\ref{eq:minimalSlip}) and (\ref{eq:parametrization}). All
colloids are oriented along the normal to the wall, such that $\boldsymbol{p}_{i}=-\hat{\boldsymbol{z}}$.
In experiment, the active force $\sim6\pi\eta bv_{s}$, is of the
order $10^{-13}$N in \citep{jiang2010active,palacci2013living},
while $10^{-11}$N in \citep{petroff2015fast}. Thus, it is orders
of magnitude larger than the Brownian force $k_{B}T/b\sim10^{-15}$N,
which we ignore for this example.

The code and snapshots from the simulations are shown in Fig.(\ref{fig:4}).
The initial random distribution of positions is provided by $\mathtt{pystokes.utils.initialConditionRandom(Np)}$.
The rigid body motion class is instantiated as $\mathtt{pystokes.wallBounded.Rbm}$,
while the Lennard-Jones inter-particle and particle-wall forces are
obtained from classes $\mathtt{pystokes.forceFields.Forces}$. A short
function computes the velocity that is passed to a standard Python
integrator which advances the system in time and saves positional
data. This is used to plot the snapshots at specified time instants. 

\section{Example 5 - Irreducible autophoretic fields}

\begin{figure}[h]
\begin{lstlisting}[language=Python,basicstyle={\scriptsize\ttfamily},breaklines=true]
# ex5.py: chemical field of an autphoretic colloid - unbounded domain (first row) and near a wall (second)
import pystokes, numpy as np, matplotlib.pyplot as plt

# particle radius, fluid viscosity, and number of particles 
b, D, Np = 1.0, 1.0/6.0, 1

#initialise  
r, p = np.array([0.0, 0.0, 5]), np.array([0.0, 0.0, 1]) 
# irreducible coeffcients  
J0 = np.ones(Np);  
J1 = pystokes.utils.irreducibleTensors(1, p) 
J2 = pystokes.utils.irreducibleTensors(2, p)

# space dimension , extent , discretization 
dim, L, Ng = 3, 10, 64; 

# instantiate the phoretic field class class 
ufield = pystokes.phoreticUnbounded.Field(radius=b, particles=Np, phoreticConstant=D, gridpoints=Ng*Ng) 
wfield = pystokes.phoreticWallBounded.Field(radius=b, particles=Np, phoreticConstant=D, gridpoints=Ng*Ng)

# plot using subplots on a given grid
rr, vv = pystokes.utils.gridYZ(dim, L, Ng)

plt.figure(figsize=(24, 8), edgecolor='gray', linewidth=4)
plt.subplot(231);  vv=vv*0;  ufield.phoreticField0(vv, rr, r, J0)   
pystokes.utils.plotContoursYZ(vv, rr, r,offset=1e-8,  title='m=0') 
plt.subplot(232);  vv=vv*0;  ufield.phoreticField1(vv, rr, r, J1)   
pystokes.utils.plotContoursYZ(vv, rr, r, offset=1e-1,title='m=1') 
plt.subplot(233);  vv=vv*0;  ufield.phoreticField2(vv, rr, r, J2)
pystokes.utils.plotContoursYZ(vv, rr, r,offset=1e2, title='Jm=2') 
plt.subplot(234);  vv=vv*0;  wfield.phoreticField0(vv, rr, r, J0)   
pystokes.utils.plotContoursYZsurf(vv, rr, r, offset=1e-8,) 
plt.subplot(235);  vv=vv*0;  wfield.phoreticField1(vv, rr, r, J1)   
pystokes.utils.plotContoursYZsurf(vv, rr, r, offset=1e-2) 
plt.subplot(236);  vv=vv*0;  wfield.phoreticField2(vv, rr, r, J2)
pystokes.utils.plotContoursYZsurf(vv, rr, r, offset=1e+2) 
\end{lstlisting}
\includegraphics[width=1\textwidth]{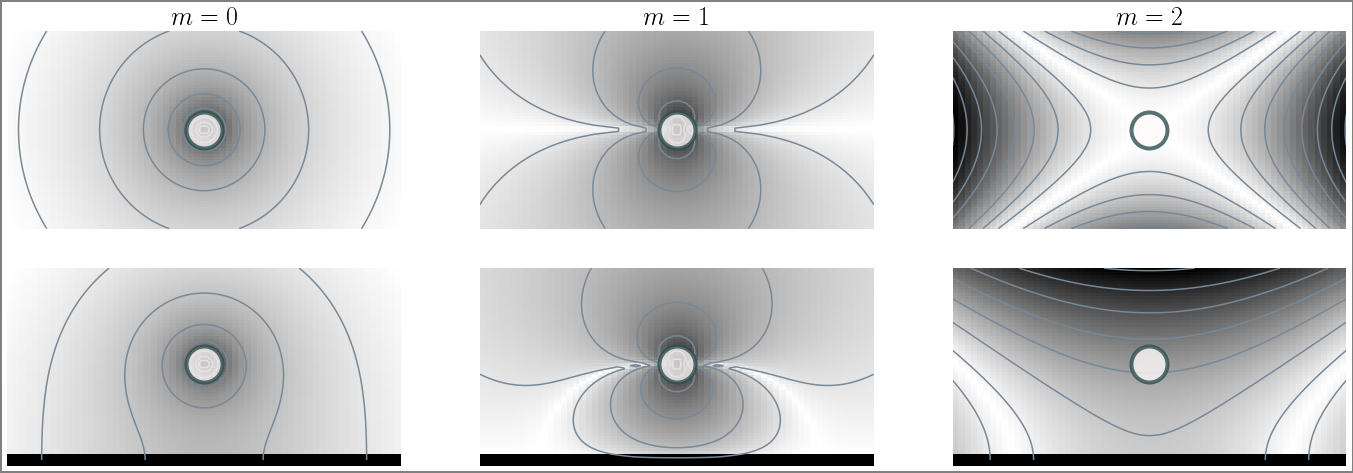}\caption{Irreducible phoretic fields: Pseudocolor and contour plot of the phoretic
field. The top is the concentration in unbounded domain, while bottom
is near a plane wall. The first panel is the concentration due to
$m=0$ mode of the flux, while second is for $m=1$, abd third is
for $m=2$. The symmetry of the profile follows that of the mode in
an unbounded domain. This symmetry is broken in $z-$direction by
the introduction of the plane wall.\label{fig:5}}
\end{figure}
Our fifth example introduces phoretic fields and, as in Example 1,
shows how to plot the irreducible parts of an active phoretic field
around a spherical particle. We take the reader step by step from
the governing PDE to the expression for the irreducible phoretic fields
that the library evaluates and plots. The notation for the particle
size, location and orientation are identical to Example 1. We encourage
the reader to note the similarities and differences between the two
examples. 
\begin{enumerate}
\item \emph{Elliptic PDE}\textbf{\emph{.}} The phoretic field $c\,(\boldsymbol{r})$
satisfies the Laplace equation in the region $V$ exterior to a sphere
and satisfies the flux boundary condition on $S$
\begin{equation}
\lim_{\boldsymbol{r}\rightarrow S}\hat{\boldsymbol{\rho}}\cdot(D\boldsymbol{\nabla}c)=-j^{\mathcal{A}}(\boldsymbol{\rho}),
\end{equation}
 where $j^{\mathcal{A}}(\boldsymbol{\rho})$ is the active flux and
$D$ is the diffusivity. The fundamental solution of the Laplace equation
is by
\begin{gather}
\nabla^{2}H=-\delta\left(\boldsymbol{r}-\boldsymbol{r'}\right)/D,
\end{gather}
where $H$ is the Green's function which may need to satisfy additional
boundary conditions. The normal derivative of the Green's function
is $L=D\hat{\rho}_{\alpha}\nabla_{\alpha}H$.
\item \emph{Boundary integral. }The fundamental solution, together with
Green's identities, gives the boundary integral representation
\begin{align}
c\,(\boldsymbol{r})=\int H(\boldsymbol{r},\,\boldsymbol{R}+\mathbf{\boldsymbol{\rho}})\,j^{\mathcal{A}}(\boldsymbol{\rho})\,\text{d}S+\int L(\boldsymbol{r},\,\boldsymbol{R}+\mathbf{\boldsymbol{\rho}})\,c(\boldsymbol{\rho})\,\text{d}S,\label{eq:bieLaplace}
\end{align}
which expresses the phoretic field in $V$ in terms of a ``single-layer''
integral involving active flux and a ``double-layer'' integral involving
the boundary phoretic field. The former is specified by the boundary
condition; the latter must be determined in terms of it. 
\item \emph{Spectral expansion. }The analytical evaluation of the two integrals
is possible if the phoretic field and the flux are expanded spectrally
in terms of tensorial spherical harmonics,
\begin{gather}
c(\boldsymbol{\rho})=\sum_{m=0}^{\infty}w_{m}\mathbf{C}^{(m)}\cdot\mathbf{Y}^{(m)}(\bm{\hat{\rho}}),\qquad j^{\mathcal{A}}(\boldsymbol{\rho})=\sum_{m=0}^{\infty}\tilde{w}_{m}\mathbf{J}^{(m)}\cdot\mathbf{Y}^{(m)}(\bm{\hat{\rho}}),
\end{gather}
where $\mathbf{C}^{(m)}$ and $\mathbf{J}^{(m)}$ are $l$-th rank
tensorial\textrm{ coefficients, symmetric and irreducible in all their
indices. The orthogonality of the tensorial harmonics implies that}
\begin{gather}
\mathbf{C}^{(m)}=\tilde{w}_{m}\int c(\boldsymbol{R}+\bm{\rho})\mathbf{Y}^{(m)}(\bm{\hat{\rho}})d\text{S},\qquad\mathbf{J}^{(m)}=w_{m}\int j(\boldsymbol{R}+\bm{\rho})\mathbf{Y}^{(m)}(\bm{\hat{\rho}})d\text{S}.
\end{gather}
\item \emph{Ritz-Galerkin discretization. }Letting the point $\boldsymbol{r}$
approach $S$ from $V$ and requiring the phoretic field to attain
its value on the boundary leads to an integral equation for it. Multiplying
both sides of the integral equation by the $l$-th harmonic and integrating
yields an infinite-dimensional system of linear equations for the
coefficients of the phoretic field,
\begin{alignat}{1}
 & \tfrac{1}{2}\mathbf{C}^{(l)}=\boldsymbol{H}^{(l,l')}\cdot\mathbf{J}^{(l')}+\boldsymbol{L}^{(l,l')}\cdot\mathbf{C}^{(l')},\label{eq:linear-system-1}
\end{alignat}
where the matrix elements $\boldsymbol{H}^{(l,\,l')}$ and $\boldsymbol{L}^{(l,\,l')}$
are give in terms of the Green's function and its normal derivative.
These can be evaluated analytically \citep{singh2019competing}.
\item \emph{Truncation: }The infinite-dimensional linear system has to be
truncated to a finite-dimensional linear system for tractability.
We truncate the system at $l=2$, so that the phoretic field and its
flux are
\begin{gather}
c(\bm{\rho})=C^{(0)}+\mathbf{C}^{(1)}\cdot\mathbf{Y}^{(1)}+\tfrac{1}{6}\mathbf{C}^{(2)}\cdot\mathbf{Y}^{(2)},\qquad4\pi b^{2}\,j^{\mathcal{A}}(\bm{\rho})=J^{(0)}+3\mathbf{J}^{(1)}\cdot\mathbf{Y}^{(1)}+5\mathbf{J}^{(2)}\cdot\mathbf{Y}^{(2)}.\label{eq:slip-truncation-1-2}
\end{gather}
The solution of the finite-dimensional linear system yields a linear
relation between the coefficients of the field and its flux, the ``elastance
relations'', $\mathbf{C}^{(m)}=-\boldsymbol{\varepsilon}^{(m,m')}\cdot\mathbf{J}^{(m')}$.
In an unbounded domain, the elastance tensors $\boldsymbol{\varepsilon}^{(m,m')}$
take on a particularly simple form: they are diagonal in the $m$
indices, $\boldsymbol{\varepsilon}^{(m,m')}\equiv\delta_{mm'}\Delta^{(m)}/4\pi Dw_{m}$
so that the single scalar $4\pi Dw_{m}$ determines them. For $m=0$
this is the familiar coefficient $1/4\pi b$ for the inverse capacitance
of a spherical conductor. 
\end{enumerate}
Inserting the truncated spectral expansions for the phoretic field
and the active flux, eliminating the unknown phoretic coefficients
for the known flux coefficients, expanding the Green's function about
the center of the sphere and finally using orthogonality of the tensorial
harmonics, we obtain the phoretic field due to each irreducible flux
mode as

\begin{gather}
c^{0}(\boldsymbol{r})=HJ^{(0)},\qquad c^{1}(\boldsymbol{r})=(\boldsymbol{\nabla}H)\cdot\mathbf{J}^{(1)},\qquad c^{2}(\boldsymbol{r})=\tfrac{4}{3}(\boldsymbol{\nabla}\boldsymbol{\nabla}H)\cdot\mathbf{J}^{(2)},
\end{gather}
We emphasise that these expressions are valid for any Green's function
of the Laplace equation, provided they satisfy the additional boundary
conditions that may be imposed. For this example we consider the Green's
function in an unbounded domain, where the flux vanishes at infinity,

\begin{equation}
H_{\alpha\beta}^{\text{o}}(\boldsymbol{r}-\boldsymbol{r}^{\prime})=\frac{1}{8\pi D}\nabla^{2}|\boldsymbol{r}-\boldsymbol{r}^{\prime}|.\label{eq:laplaceUnbounded}
\end{equation}
and in a domain confined by an infinite planar wall at which the flux
vanishes,
\begin{equation}
H_{\alpha\beta}^{\text{w}}(\boldsymbol{r},\boldsymbol{r}^{\prime})=H_{\alpha\beta}^{\text{o}}(\boldsymbol{r}-\boldsymbol{r}^{\prime})+H_{\alpha\beta}^{\text{o}}(\boldsymbol{r}-\boldsymbol{r}^{*\prime}).\label{eq:wallGLaplace}
\end{equation}
Here, as earlier, $\boldsymbol{r}^{*\prime}=\boldsymbol{\mathcal{M}}\cdot\boldsymbol{r}'$
is the mirror image of the point, where\textrm{ $\boldsymbol{\mathcal{M}}=\boldsymbol{I}-2\mathbf{\hat{z}}\mathbf{\hat{z}}$}
is the reflection operator. 

The code listed in Fig.(\ref{fig:5}) computes the irreducible concentration
profile for $m=0$,1,2 modes of the active flux for radius $b=1$,
diffusion constant $D=1$, location $\boldsymbol{R}=(0,0,5)$ and
orientation $\boldsymbol{p}=(0,0,1).$ The coefficients are parametrised
as $J^{(0)}=1$, $J_{\alpha}^{(1)}=p_{\alpha}$ and $J_{\alpha\beta}^{(2)}=p_{\alpha}p_{\alpha}-\delta_{\alpha\beta}/3$.
These are supplied to the $\mathtt{ufield}$ class which is instantiated
for an unbounded domain. The $m$-th irreducible component of the
field is computed by the calling the function $\mathtt{ufield.phoreticFieldm}$.
This is passed to a generic plotting function to compute the pseudocolor
plot in a plane of symmetry. These are shown in the code output where
the spherical , polar, and nematic symmetries of the $m=0$, 1 and
$2$ modes can be seen. The second row shows the same fields near
a plane wall, obtained by changing the flow instantiation to $\mathtt{wfield}$
class. The fields have reduced symmetry in $z-$direction due to the
introduction of the plane wall. 

\section{Example 6 - Autophoretic arrest of flow-induced phase separation\label{sec:Phoretic-arrest-of}}

\begin{figure}
\begin{lstlisting}[language=Python,basicstyle={\scriptsize\ttfamily},breaklines=true]
# ex6.py: Arrested clustering of autphoretic colloids near a wall

import pystokes, numpy as np, matplotlib.pyplot as plt
# particle radius, self-propulsion speed, number and fluid viscosity 
b, vs, Np, eta = 1.0, 1.0, 256, 0.1

#initialise 
r = pystokes.utils.initialCondition(Np)  # initial random distribution of positions 
p = np.zeros(3*Np); p[2*Np:3*Np] = -1    # initial orientation of the colloids

rbm      = pystokes.wallBounded.Rbm(radius=b, particles=Np, viscosity=eta) 
phoresis = pystokes.phoreticWallBounded.Phoresis(radius=b, particles=Np, phoreticConstant=eta) 
forces   = pystokes.forceFields.Forces(particles=Np) 

N1, N2 = 3*Np, 6*Np  # define two constants for convenience
def rhs(rp):     
	"""
	* right hand side of the rigid body motion equations     
	* rp: is the array of position (r) and orientations (p) of the colloids     	
	* returns \dot{rp} so that rp can be updated using an integrator     
	"""     	
	#initialise the positions and orientation, forces at each time step
	r = rp[0:N1];   p = rp[N1:N2];        
	F,v,o,C1 = np.zeros(N1),np.zeros(N1),np.zeros(N1),np.zeros(N1)          
	
	# rbm contributions from body forces          
	forces.lennardJonesWall(F, r, lje=0.012, ljr=5, wlje=1.2, wljr=3.4)     
  rbm.mobilityTT(v, r, F)        

	#phoretic field on the surface of colloids     
	J0, J1 = .4*np.ones(Np), pystokes.utils.irreducibleTensors(1, p)  
	phoresis.elastance10(C1, r, J0);  phoresis.elastance11(C1, r, J1)          

	# active contributions to the rbm     
	M0=1;  V1s=-pystokes.utils.couplingTensors(0, p, M0)*C1;  V3t=0.6*V1s     
	rbm.propulsionT3t(v, r, V3t); v += V1s     
	return np.concatenate( (v,o) )

# simulate the resulting system and plot at specific time instants 
Tf, Npts = 300, 256 
pystokes.utils.simulate(np.concatenate((r,p)), Tf,Npts,rhs,integrator='odeint', filename='arrestedClusters')
pystokes.utils.plotConfigs(t=[1, 100, 200, 256], ms=60, tau=(Tf/Npts)/(b/vs), filename='arrestedClusters')
\end{lstlisting}
\includegraphics[width=1\textwidth]{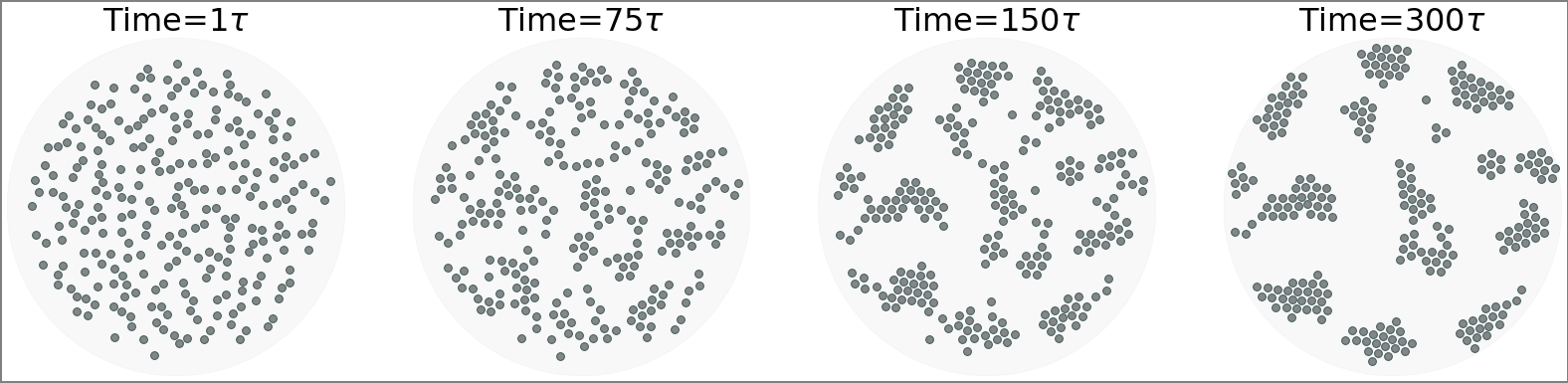}\caption{Phoretic arrest of flow-induced phase separation. In this example,
we extend the results of Fig.(\ref{fig:4}) by adding phoretic interactions.
We show that this can lead to arrest of flow-induced phase separation
when the phoretic interactions between the colloids are repulsive.
The length scale at which the clusters are arrested is proportional
the self-propulsion speed of an isolated colloid \citep{singh2019competing},
in excellent agreement with experimental observations \citep{theurkauff2012dynamic,buttinoni2013DynamicClustering}.
\label{fig:6} }
\end{figure}
Our sixth example describes the modification of active slip by phoretic
interactions between active particles and the resulting arrest of
flow-induced phase separation seen in Example 4. The slip, from being
an apriori specified quantity for each particle, must now be computed
from the interacting phoretic fields of all particles. In the previous
example, we obtained the phoretic field for a specified surface flux.
Here we show how to coupled it to the Stokes equation and obtain both
the hydrodynamic and phoretic interactions of active colloids. The
coupling of the Stokes and Laplace equation and Laplace equations
is through the following expression for the active slip \citep{anderson1989colloid},
\begin{alignat}{1}
\boldsymbol{v}^{\mathcal{A}}(\boldsymbol{\rho}_{i}) & =\mu_{c}(\boldsymbol{\rho}_{i})\boldsymbol{\nabla}_{s}\,c(\boldsymbol{\rho}_{i}).\label{eq:slipConcGrad}
\end{alignat}
In the previous section, we showed that $\mathbf{C}_{i}^{(m)}=-\boldsymbol{\varepsilon}{}_{ik}^{(m,m')}\cdot\mathbf{J}_{k}^{(m')}.$
This solution can be used to obtain the coefficients of the slip as
$\mathbf{V}_{i}^{(l)}=-\boldsymbol{\chi}^{(l,m)}\cdot\mathbf{C}_{i}^{(m)},$
where $\boldsymbol{\chi}^{(l,m)}$ is a coupling tensor of rank $(l+m)$
that depends on the phoretic mobility $\mu_{c}$ \citep{singh2019competing}.
Thus the problem is fully specified once the coefficients of the flux
and phoretic mobility on the surface of all the particles is given. 

For this example, we consider an active surface flux $j^{\mathcal{A}}$
with spherical and polar modes and a phoretic mobility that is constant,
\begin{equation}
4\pi b^{2}\,j^{\mathcal{A}}(\boldsymbol{\rho}_{i})=J^{(0)}+3\,J^{(1)}\boldsymbol{p}_{i}\cdot\boldsymbol{\hat{\rho}}_{i},\qquad4\pi b^{2}\,\mu_{c}(\boldsymbol{\rho}_{i})=M^{(0)}.\label{eq:minimal}
\end{equation}
Phoretic interactions are computed with he no-flux condition at the
plane wall using the Green's function of Eq.(\ref{eq:wallGLaplace}).
Hydrodynamic interactions are computed with the no-slip on the plane
wall using the Lorentz-Blake tensor of Eq.(\ref{eq:stokesGwall}). 

The code and snapshots from the simulations are shown in Fig.(\ref{fig:6}).
The $\mathtt{pystokes.phoreticWallBounded.Phoresis}$ class is used to compute
the active slip which is fed into $\mathtt{pystokes.wallBounded.Rbm}$
to compute the rigid body motion. As before, a short function is used
to compute the velocity that is passed on to a standard Python integrator
which advances the system in time and saves position data. This is
used to plot the snapshots at specified time instants.
\begin{figure}[H]
\begin{lstlisting}[language=Python,basicstyle={\scriptsize\ttfamily},breaklines=true]
# ex7.py: Benchmarks
import numpy as np, matplotlib.pyplot as plt, pystokes, time, matplotlib as mpl

#computes time taken to simulate Np particles
def timeTaken(Np):     
	b, vs, eta = 1.0, 1.0, 0.1;    r = 2*np.linspace(-3*Np, 3*Np, 3*Np)     
	p, v       = np.ones(3*Np), np.zeros(3*Np)     
	RBM = pystokes.wallBounded.Rbm(radius=b, particles=Np, viscosity=eta)     
	V3t = p;    t1 = time.perf_counter();    RBM.propulsionT3t(v, r, p)     
	return time.perf_counter() - t1

xP  = np.arange(2000, 21000, 2000) 
tP1 = np.zeros(np.size(xP))

for i in range(np.size(xP)):     
	tP1[i]=timeTaken(xP[i])

plt.figure(figsize=(16, 14))
mpl.rc('hatch', color='k', linewidth=22.5) 
mss=24; plt.xticks(fontsize=mss);   plt.yticks(fontsize=mss) 
plt.semilogy(xP, tP1, '-*', ms=mss+3, label="1 Core", lw=3, color='lightslategray', mfc='w', mew=2, alpha=1) 
plt.legend(fontsize=mss, loc=4) 
plt.xlabel('# colloids (thousands)', fontsize=mss+5) 
plt.ylabel('CPU time (secs)', fontsize=mss+5) 
plt.grid()

# repeat for different number of cores. We plot a precompiled set of benchmarks
\end{lstlisting}
\includegraphics[width=1\textwidth]{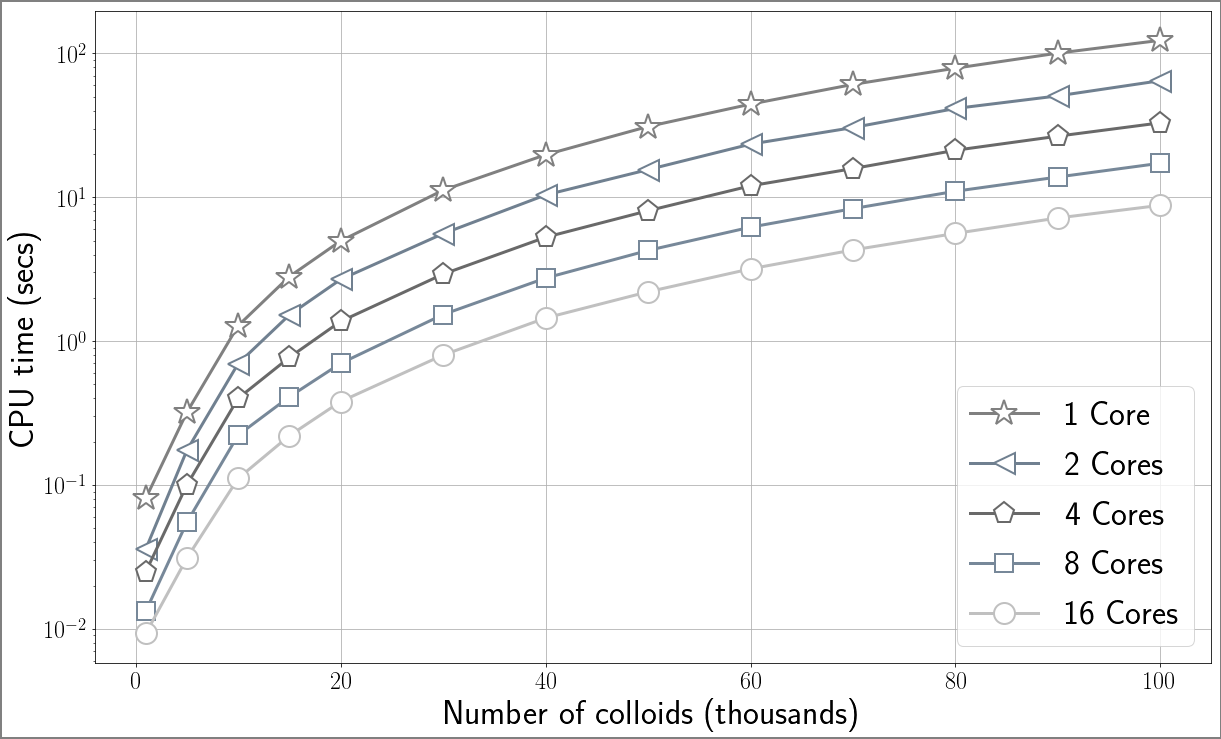}\caption{Benchmarks for computing the rigid body motion due to $\mathbf{V}^{(3t)}$
slip mode in the unbounded geometry of Stokes flow, on a 16-core machine,
using PyStokes. The present implementation shows a linear scaling
with the number of CPU cores, and quadratic scaling with number of
particles, when the matrix-vector products are computed as direct
sums.\label{fig:7}}
\end{figure}

\section{Code performance}

Our final example is on performance and benchmarks. We show that our
codes scales linearly with the number of CPU cores and quadratically
with $N$, the number of particles simulated. In the current implementation,
the velocities of about $10^{5}$ particles can be computed in a few
seconds for a mode of the active slip, as shown in Fig.(\ref{fig:7}).
With current many-core architectures a dynamic simulation of about
$N\sim10^{5}$ is within reach. For larger number of particles, accelerated
summation methods are desirable, which can reduce the cost to $\mathcal{O}(N\log N)$
\citep{barnes1986hierarchical,sierou2001accelerated} or even $\mathcal{O}(N)$
\citep{greengard1987fast,sangani1996n,ladd1994numericala}. These
methods can be implemented in the present numerical architecture as
an improvement over the direct kernel sum, while maintaining the overall
library structure. 

\section{What else ?}

In the six examples above we demonstrated the use of PyStokes library
for simulating hydrodynamic and phoretic phenomena. These do not exhaust
the capabilities of the library and much else can be done with them.
We conclude by listing implemented, implementable and unimplementable
features. \emph{Implemented but not shown:}\textbf{ }the library supports
periodic geometries \citep{singh2018microhydrodynamics} and parallel
plane walls \citep{thutupalli2018FIPS,sarkar2017ns}. Polymers \citep{laskar2015brownian},
membranes and other hierarchical assemblies of active particles can
be simulated.\emph{ Can be implemented:}\textbf{ }other boundary conditions,
for instance flows interior to a spherical domain such as a liquid
drop, near-field lubrication interactions, and numerical solutions
of the linear system can be implemented in the current design, with
treecode or fast-multipole accelerations.\emph{ Cannot be implemented:}\textbf{
}irregular geometries for which analytical forms of the Green's functions
of the Laplace and Stokes equations cannot be evaluated analytically
and/or irregularly shaped particles on whose boundaries globally defined
spectral basis functions are not available. A shorter version of this
paper has been published in JOSS \citep{singh2020PyStokes}. 

\section*{Acknowledgement}

We would like to thank collaborators and colleagues, in both theory
and experiment, for numerous discussions that have enriched our understanding
of hydrodynamic and phoretic phenomena. In alphabetical order, they
are: Ayan Banerjee, Mike Cates, Shyam Date, Aleks Donev, Erika Eiser,
Daan Frenkel, Somdeb Ghose, Ray Goldstein, John Hinch, Abhrajit Laskar,
Tony Ladd, Raj Kumar Manna, Ignacio Pagonabarraga, Dave Pine, Thalappil
Pradeep, Rajiah Simon, Howard Stone, Ganesh Subramanian, P. B. Sunil
Kumar, and Shashi Thutupalli. We acknowledge design ideas and code
contributions from Rajeev Singh and Abhrajit Laskar in the initial
stages of development. This article is a contribution in commemoration
of Sir George Gabriel Stokes in his bicentennial \citep{stokes200}.
This work was funded in parts by the European Research Council under the EU's Horizon 2020 Program, Grant No. 740269; 
a Royal Society-SERB Newton International Fellowship to RS; and an Early Career Grant to RA from the Isaac Newton Trust.

\end{document}